\documentclass[%
 aip,
 amsmath,amssymb,
 reprint,
]{revtex4-1}

\usepackage{graphicx}
\usepackage{subfigure}
\usepackage{subcaption}
\usepackage{dcolumn}
\usepackage{bm}

\usepackage[utf8]{inputenc}
\usepackage[T1]{fontenc}
\usepackage{mathptmx}
\usepackage{etoolbox}
\usepackage[overlay,absolute]{textpos}\newcommand\PlaceText[3]{\begin{textblock*}{10in}(#1,#2)#3\end{textblock*}}

\makeatletter
\def\@email#1#2{%
 \endgroup
 \patchcmd{\titleblock@produce}
  {\frontmatter@RRAPformat}
  {\frontmatter@RRAPformat{\produce@RRAP{*#1\href{mailto:#2}{#2}}}\frontmatter@RRAPformat}
  {}{}
}%
\makeatother
\begin{document}

\PlaceText{12mm}{8mm}{APL Photonics \textbf{11}, 086111 (2026); https://doi.org/10.1063/5.0321309}

\title{Phase-Randomized Laser Pulse Generation at 10 GHz for Quantum Photonic Applications}

\newcommand{\equalcontrib}{These authors contributed equally to this work.}

\author{Y.~S.~Lo}
\email{yuen.lo@toshiba.eu}          
\thanks{\equalcontrib}               
\affiliation{Toshiba Europe Ltd, Cambridge, UK}

\author{A.~H.~Brzosko}
\email{ab2967@cam.ac.uk}   
\thanks{\equalcontrib}              
\affiliation{Toshiba Europe Ltd, Cambridge, UK}
\affiliation{Cambridge University Engineering Department, Cambridge, UK}

\author{P.~R.~Smith}
\affiliation{Toshiba Europe Ltd, Cambridge, UK}

\author{R.~I.~Woodward}
\affiliation{Toshiba Europe Ltd, Cambridge, UK}

\author{D.~G.~Marangon}
\affiliation{Toshiba Europe Ltd, Cambridge, UK}

\author{J.~F.~Dynes}
\affiliation{Toshiba Europe Ltd, Cambridge, UK}

\author{S. Ju\'arez}
\affiliation{Toshiba Europe Ltd, Cambridge, UK}
\affiliation{Escuela de Ingenier\'ia de Telecomunicaci\'on, University of Vigo, Vigo, Spain}

\author{T.~K.~Para\text{\"i}so}
\affiliation{Toshiba Europe Ltd, Cambridge, UK}

\author{R.~M.~Stevenson}
\affiliation{Toshiba Europe Ltd, Cambridge, UK}

\author{A.~J.~Shields}
\affiliation{Toshiba Europe Ltd, Cambridge, UK}


\begin{abstract}
Gain-switching of laser diodes is a well-established technique for generating optical pulses with random phases, where the quantum randomness arises naturally from spontaneous emission. However, the maximum switching rate is limited by the time required for phase diffusion: at high repetition rates, residual photons in the cavity seed subsequent pulses, leading to phase correlations, which degrade randomness. We present a method to overcome this limitation by employing an external source of spontaneous emission in conjunction with the laser. Our results show that this approach substantially suppresses interpulse phase correlations, restoring phase randomization at repetition rates as high as 10 GHz. This technique opens new opportunities for high-rate quantum key distribution and quantum random number generation.
\end{abstract}

\maketitle

\section{\label{sec:level1}Introduction}

Gain switching is a widely used technique for generating optical pulses in semiconductor laser diodes, achieved by directly modulating the electrical pump power to periodically drive the laser above and below the threshold. This produces a train of short, well-defined optical pulses. Beyond its established role in classical optical communications, gain switching has also found significant applications in the field of quantum key distribution (QKD) where two parties can exchange secret keys with security guaranteed by the fundamental laws of physics \cite{ Gisin2002, bb84}. 

In the most widely used QKD protocols, a key requirement in the security proofs is that the phase of each quantum signal must be uniformly random \cite{Inamori2007, Gottesman2004, nonrandom}. Ideal single-photon sources naturally provide photons with undefined phase, but practical QKD systems rely on weak coherent pulses (WCPs) from attenuated lasers. To meet the security requirements when using WCPs, including in decoy-state protocols \cite{Lo2005Decoy}, the global phase of every pulse must be randomized. This can be readily achieved through gain switching. During the below-threshold interval between pulses, the intracavity field decays and the emission is dominated by spontaneous emission, in which the optical phase evolves stochastically and can be described by phase-diffusion models \cite{Abellan2014, Septriani2020, Quirce2021}. When the laser is subsequently driven above threshold, stimulated emission amplifies an intracavity field whose phase has already been randomized, thus each pulse emerges with a global phase that is effectively random and uncorrelated with previous pulses. Due to its simplicity and effectiveness, this method is adopted in many practical QKD systems \cite{Yuan2018, Boaron2018, Woodward2021a, Agnesi2020}.

Spontaneous emission provides an important resource for quantum randomness and underpins the operation of many quantum random number generators (QRNGs) \cite{Marangon2024, Abellan2014, Yuan2014.2, Alkhazragi2023, HerreroCollantes2017}. To extract the phase randomness in the gain-switched pulses, an asymmetric Mach–Zehnder interferometer (AMZI) can be used to interfere consecutive pulses, thereby converting random phase variations into directly measurable intensity fluctuations.

One crucial requirement for this technique is that the laser must remain below threshold long enough for the carrier population to deplete sufficiently, such that the contribution of stimulated emission photons to the subsequent pulse generation is negligible. If residual photons from the previous emission persist in the cavity, they seed the next pulse and imprint their phase onto it, thereby introducing unwanted correlations between pulses and degrading the practical security of QKD \cite{CurrasLorenzo2024ImperfectPhase}. While practical implementations are subject to electronic bandwidth constraints, this condition constrains the maximum clock rate of QKD systems—and consequently the maximum achievable key rate. In state-of-the-art QKD implementations, clock rates are therefore typically limited to only a few gigahertz  \cite{Li2023, Grunefelder2023, Grunenfelder2020, Yuan2018}.

The same limitation also applies to phase-noise-based QRNGs \cite{Abellan2014, Xu2012, Yuan2014, Marangon2024}. Although phase diffusion in these sources has been studied extensively, the randomness ultimately relies on the laser's intrinsic spontaneous emission to randomize the phase within the below-threshold interval. As the repetition rate increases, this interval shortens, until the intrinsic spontaneous emission can no longer fully randomize the phase before the next pulse forms; strong phase correlations have accordingly been reported when the laser is gain-switched at rates as high as 10 GHz \cite{Yuan2014}.  

Here we introduce, to the best of our knowledge for the first time, a technique that restores phase randomization at high pulse repetition rates by utilizing an external source of spontaneous emission. This enables phase randomization at an unprecedented pulse repetition rate of 10 GHz, overcoming the clock-rate limitation imposed by interpulse phase correlations. This provides a basis for both higher-rate phase-noise QRNG and higher-clock-rate QKD transmitters.

\section{Phase randomization with externally injected spontaneous emission photons}

\begin{figure*}[tbph!]
	\includegraphics[width=406.33pt,height=249.32625pt]{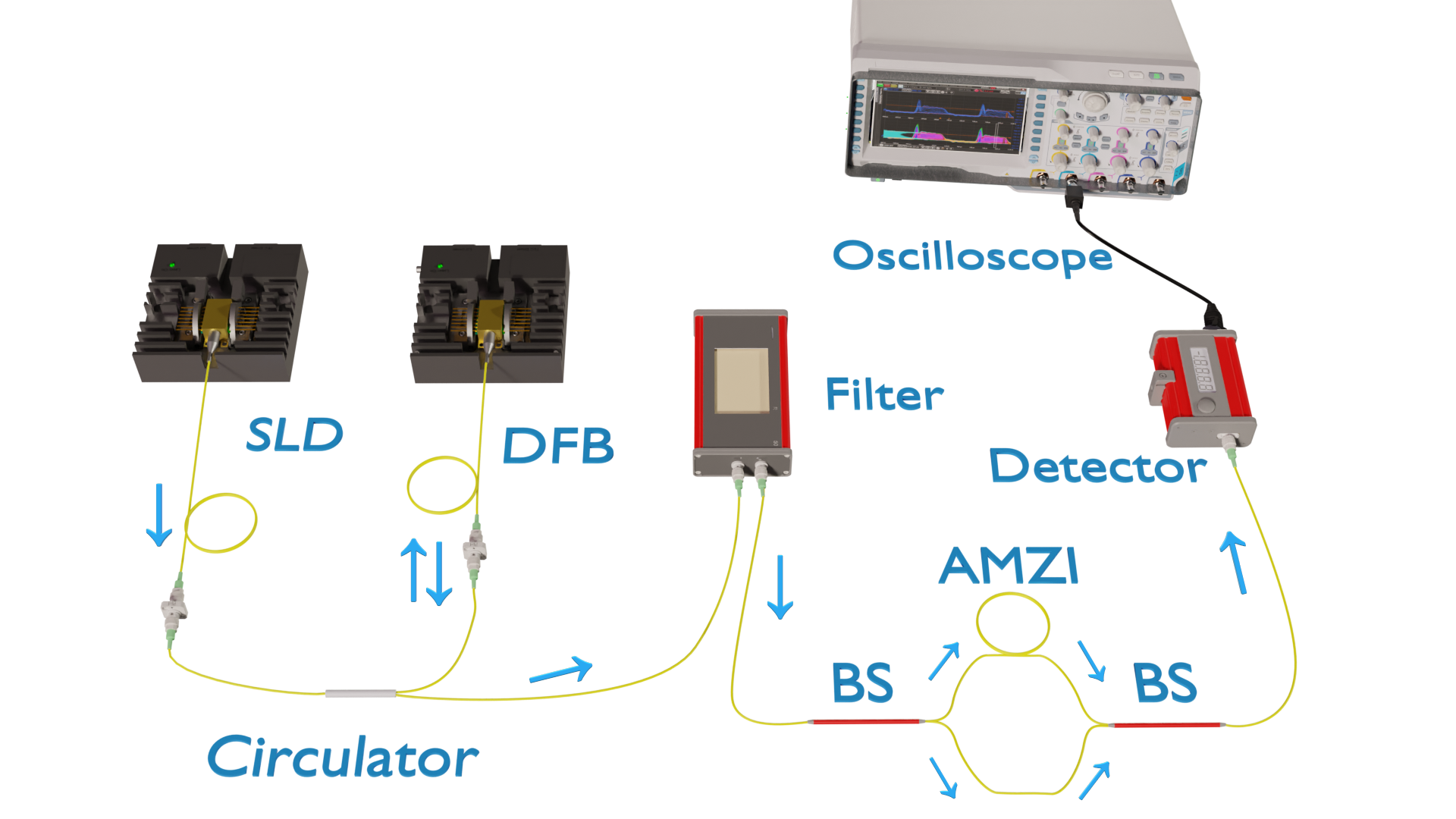}
	\caption{Experimental setup for phase randomization achieved via externally injected ASE photons. SLD, Superluminescent Diode; DFB, Distributed Feedback Laser; AMZI, Asymmetric Mach-Zehnder Interferometer; RF, Radio Frequency Driving Signal (1-10 GHz); BS, 50:50 beam splitter.}
	\label{setup}
\end{figure*}

In a gain-switched semiconductor laser, the optical gain is periodically driven above and below the lasing threshold. During the above-threshold portion of each cycle, laser oscillation is sustained by stimulated emission, leading to the build-up of the intracavity field whose phase is well defined and evolves steadily in time. In contrast, during the below-threshold interval the coherent intracavity field rapidly decays, and the optical field is dominated by spontaneous emission. In this regime, spontaneously emitted photons couple into the cavity mode with random phases, introducing stochastic perturbations to the optical field. As a result, in the course of successive spontaneous emission events, the phase of the optical field undergoes a diffusive random walk, commonly referred to as phase diffusion. This fluctuation in phase gives rise to a fundamental broadening of the optical spectrum, described by the Schawlow–Townes linewidth. For operation below threshold, the linewidth can be expressed as \cite{Henry1986}

\begin{equation}
   \Delta\nu = \frac{R_{\mathrm{sp}}}{2\pi S}  
\end{equation}

\noindent where \( R_{\mathrm{sp}} \) denotes the rate at which spontaneous emission photons are coupled into the cavity mode and \( S \) is the intracavity photon number associated with that mode. This linewidth corresponds to a phase-diffusion constant
\begin{equation}
   D_{\phi} = \pi \Delta\nu 
\end{equation}

\noindent which leads to a phase-variance of
\begin{equation}
\langle (\Delta\phi)^2 \rangle = 2 D_{\phi} t = 2\pi \Delta\nu t    
\end{equation}

Consequently, the degree of phase randomization in each optical pulse is determined by the rate of phase diffusion during the below-threshold interval and the corresponding duration. From Eq.~(3), it is evident that a larger linewidth corresponds to a faster rate of phase diffusion within a time interval $t$. In a gain-switched laser operating at high clock rates, the interval between successive pulses can be shorter than the characteristic phase diffusion time required for effective phase randomization, leading to phase correlations between pulses.

To overcome this limitation, we propose to inject broadband amplified spontaneous emission (ASE) light into the laser cavity. The externally injected ASE photons effectively augment the rate of stochastic perturbations, resulting in an effective spontaneous emission rate \( R_{\mathrm{sp,eff}} = R_{\mathrm{sp}} + R_{\mathrm{ASE}}\). This manifests as a broadened linewidth which translates into a larger phase diffusion constant \(D_{\phi}\), thereby accelerating the phase diffusion process during the below-threshold period. Consequently, the phase variance $\langle (\Delta\phi)^2 \rangle$ accumulated during the below-threshold period becomes much larger, allowing each gain-switched pulse to be initiated from a statistically random and independent phase even at multi-gigahertz clock rates. This preserves the required phase randomness and alleviates the time constraint on the pulse repetition rate. We note that this description is a simplified heuristic picture, aimed at giving the physical intuition behind the mechanism of the phase diffusion acceleration during the below-threshold interval. A complete stochastic field-equation model is outside the scope of this paper.

To practically implement this ASE injection, we utilize a superluminescent diode (SLD) as our source of amplified spontaneous emission. To elucidate this, it is worth briefly comparing an SLD with a light emitting diode (LED) and a laser diode in terms of their operating mechanism and output properties. LEDs, lacking a gain medium or feedback mechanism, emit light through spontaneous emission, resulting in a broad spectrum and non-directional output. SLDs, on the other hand, incorporate a gain medium within a waveguide, allowing spontaneous emission to be amplified as it makes a single pass through the gain medium. To prevent optical feedback, which would lead to lasing, the waveguide is often tilted \cite{Alphonse1988, Kafar2015} and coated with anti-reflective material \cite{Kaminow1983}, deliberately leaking any reflections outside the waveguide. As a result, SLDs produce
amplified spontaneous emission (ASE), which has a higher and more directional optical output than LED light, yet without the narrow spectral width and coherence of a laser diode.
The laser diodes distinguish themselves by having a cavity with mirrors (or facets) that provide positive optical feedback, leading to stimulated emission. This process yields a
highly coherent, narrowband, and directional beam of light.

As the underlying physical process in SLD is based on spontaneous emission, which
is inherently quantum in nature, SLDs have long been used as a source for QRNGs \cite{Li2011, Wei2012, Wei2017, Guo2021}. However, all SLD-based QRNGs reported to date extract quantum randomness from intensity fluctuation in the SLD output. This represents a key distinction from our approach, in which we exploit the randomness in the \textit{phase} of the ASE photons. This is achieved by injecting the ASE photons into the cavity of a laser diode which subsequently seeds the generation of phase randomized coherent optical pulses. Consequently, our scheme enables not only higher-rate phase-noise-based QRNGs, but also provides a basis for higher-rate QKD transmitters.

\section{Results}

In this section, we present the results of a series of experiments to study the effect of the external injection of ASE photons to the gain-switched laser. The experimental setup is shown in Fig. \ref{setup}. An SLD is coupled to the cavity of a distributed feedback (DFB) laser
through a circulator. The SLD has a built-in isolator and has a central wavelength of 1550 nm with a 3 dB bandwidth of 33 nm. The DFB laser has a central wavelength at 1547 nm, with a modulation bandwidth of 18 GHz. The laser is driven by an electrical waveform from a signal generator (with a frequency range of 9 kHz to 40 GHz), combined with a DC bias through a bias-tee. The SLD is operated in CW mode and the injection power can be adjusted through its DC bias. A bandpass spectral filter is used to reduce the spectral noise. An AMZI with delays matching the interval of the optical pulses is used to measure the relative phases of the optical pulses. The output is detected by a 40 GHz photodiode in the linear regime and recorded on a fast oscilloscope with a sampling rate of 80 GSample/s, a bandwidth of 33 GHz, and an internal 8-bit analog-to-digital converter. The high sampling rate allows accurate reconstruction of the temporal pulse profile. The acquisition duration is chosen so that the recorded waveform contains at least 10$^6$ pulses, enabling subsampling to extract an equivalent number of samples for subsequent statistical analysis.

When interfering two coherent pulses with a constant phase difference $\Delta \phi$,
the output intensity $I$ is proportional to (1 + cos($\Delta \phi$)). When the pulses are phase-randomized, $\Delta \phi$ has a uniform distribution and the resulting intensity distribution follows an arcsine profile, exhibiting two peaks at the minimum and maximum intensities. This is a characteristic profile which distinguishes true randomness that we want to measure from noise sources, which will be unavoidably added to the output and exhibit a Gaussian profile (e.g. the electronic noise). Additionally, the signal autocorrelation is used to analyze the traces recorded by the oscilloscope to verify the absence of phase correlations.

\begin{figure*}[tbph!]
\includegraphics[scale=0.4]{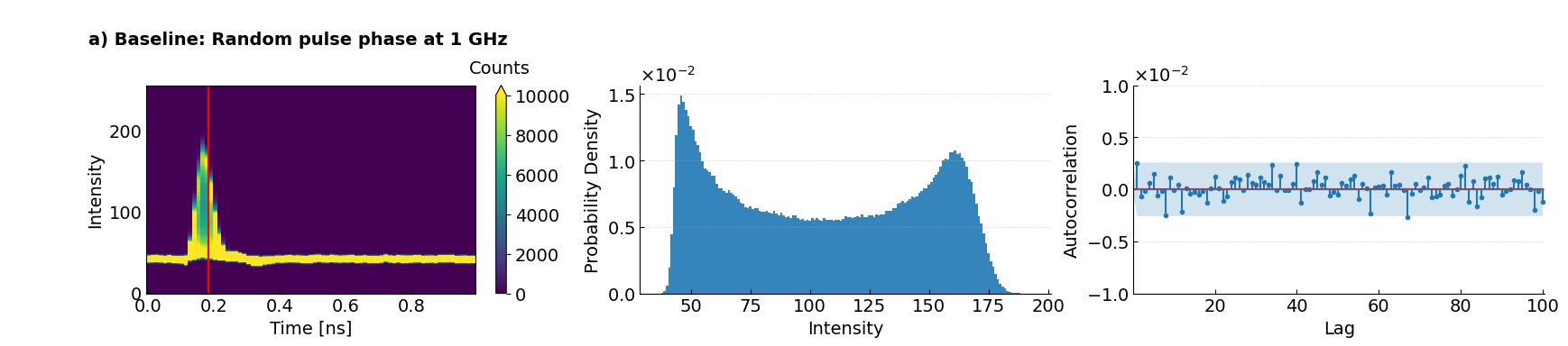}
\includegraphics[scale=0.4]{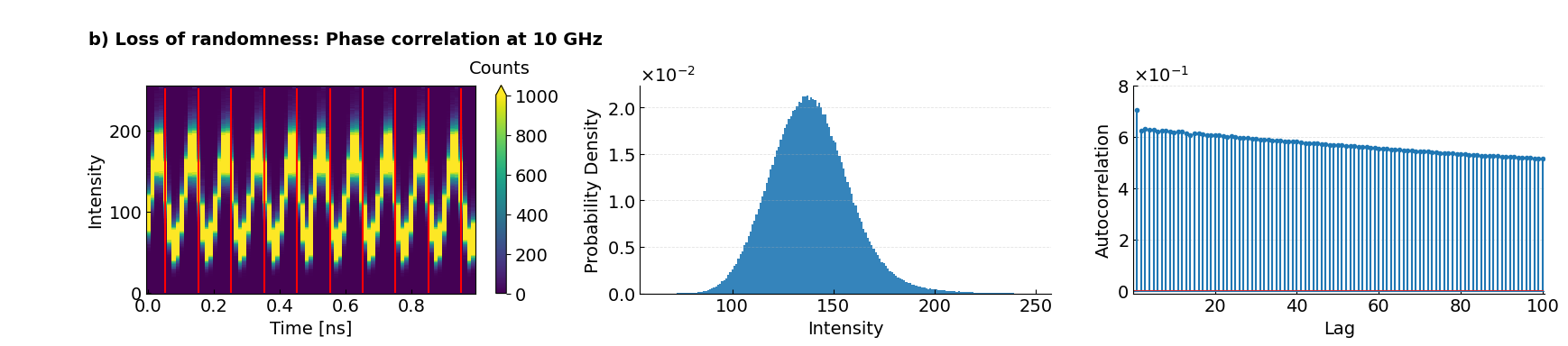}
\includegraphics[scale=0.4]{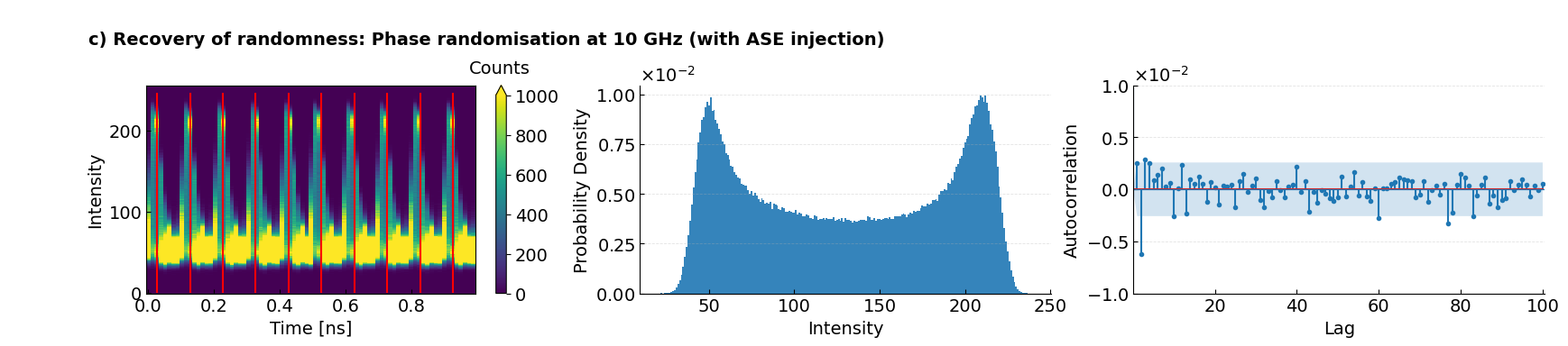}
\caption{Experimental results of the DFB laser under gain-switching at 1 GHz (a), 10 GHz (b), and 10 GHz with external ASE light injection (c). \textbf{Left column:} Density plots of the oscilloscope traces, with the sampling points indicated by red lines. \textbf{Middle column:} The corresponding intensity distributions at the sampling point (expressed as raw analog-to-digital converter (ADC) output, ranging between 0 and 255). \textbf{Right column:} Autocorrelation functions of the intensity at the sampling point, computed between samples separated by a specified number of lags. The blue shaded region denotes the 99\% confidence interval bounds and lag 0 is omitted for clarity. All analyses are performed over 10$^6$ samples.}
\label{Img:Exp10g}
\end{figure*}

We begin by gain-switching the laser at 1 GHz, a typical repetition rate that has been widely demonstrated to induce robust phase randomization in trains of pulses. As expected, the switch-off time between successive pulse generations is sufficiently long to induce seeding from spontaneous emission, characterized by an arcsine intensity distribution, with the samples being uncorrelated, as confirmed by the autocorrelation analysis (Fig. \ref{Img:Exp10g}a). The result at 1 GHz thus establishes a benchmark for our experiments.

Next, the repetition rate of the laser is increased to 10 GHz, at which point a clear departure from the desired phase-randomized behavior is observed. As shown in Fig. \ref{Img:Exp10g}b, the characteristic arcsine intensity distribution disappears, and there are strong interpulse correlations. This is evidenced by a positive autocorrelation that is at least an order of magnitude higher than in the 1 GHz case and with a slow decay, thus making the source entirely unsuitable for use in QRNG or QKD transmitter applications at this clock rate. The laser drive parameters in this case are 45.5 mA bias and 4.6 V peak-to-peak RF signal.

Upon externally injecting ASE photons into the laser cavity, the arcsine intensity distribution is immediately recovered (Fig. \ref{Img:Exp10g}c), without changing the laser's driving condition. The autocorrelation is also restored to a level comparable to the 1 GHz case. In this experiment, 19 mW of ASE was injected, with its power distributed over a 3-dB bandwidth of 33 nm. Since the laser cavity acts as a passive spectral filter, only the portion of the ASE spectrum lying within the cavity resonance can effectively seed the intracavity field. The cavity resonance width is determined by the photon lifetime $\tau_p$ via
\begin{equation}
\Delta\nu_{cav} \approx \frac{1}{2\pi \tau_p}
\end{equation}
If we consider a typical $\tau_p$ $\approx$ 4.5 ps \cite{Lovic2021PhaseNoiseQRNG}, this gives $\Delta\nu_{cav}$ $\approx$ 35 GHz, corresponding to about 0.28 nm at 1550 nm. Therefore, only about 0.16 mW of the injected ASE is spectrally capable of coupling into the lasing mode (ignoring all other coupling losses). This power is in fact comparable to the injection powers commonly used for optical injection locking \cite{Rosado2018}. We repeat the experiment at 5 GHz and 8 GHz (see Appendix), confirming the general applicability of our technique.

\begin{figure*}[tbph!]
  \centering
  \subfigure[ ]{\includegraphics[width=249pt]{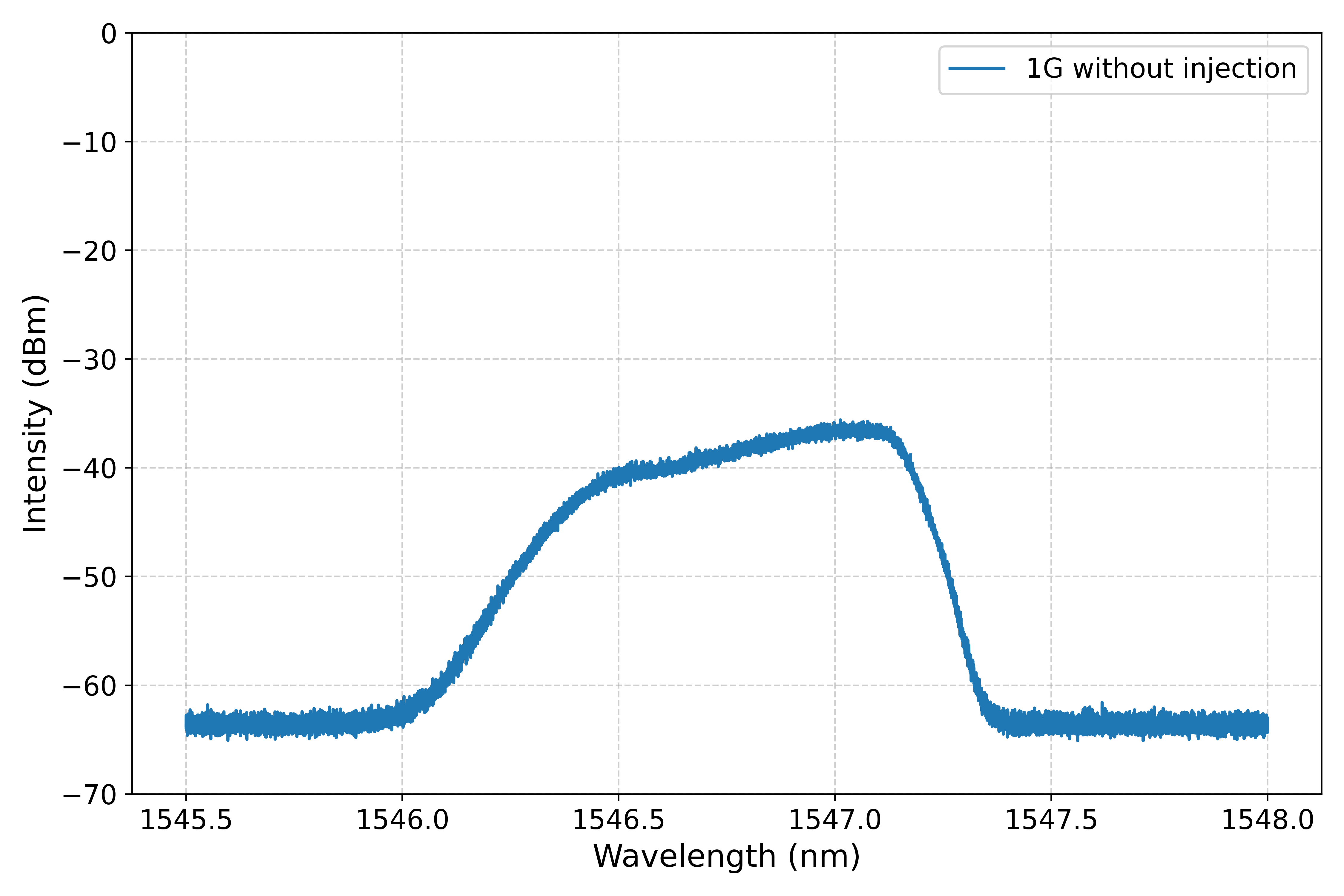}} 
  \subfigure[ ]{\includegraphics[width=249pt]{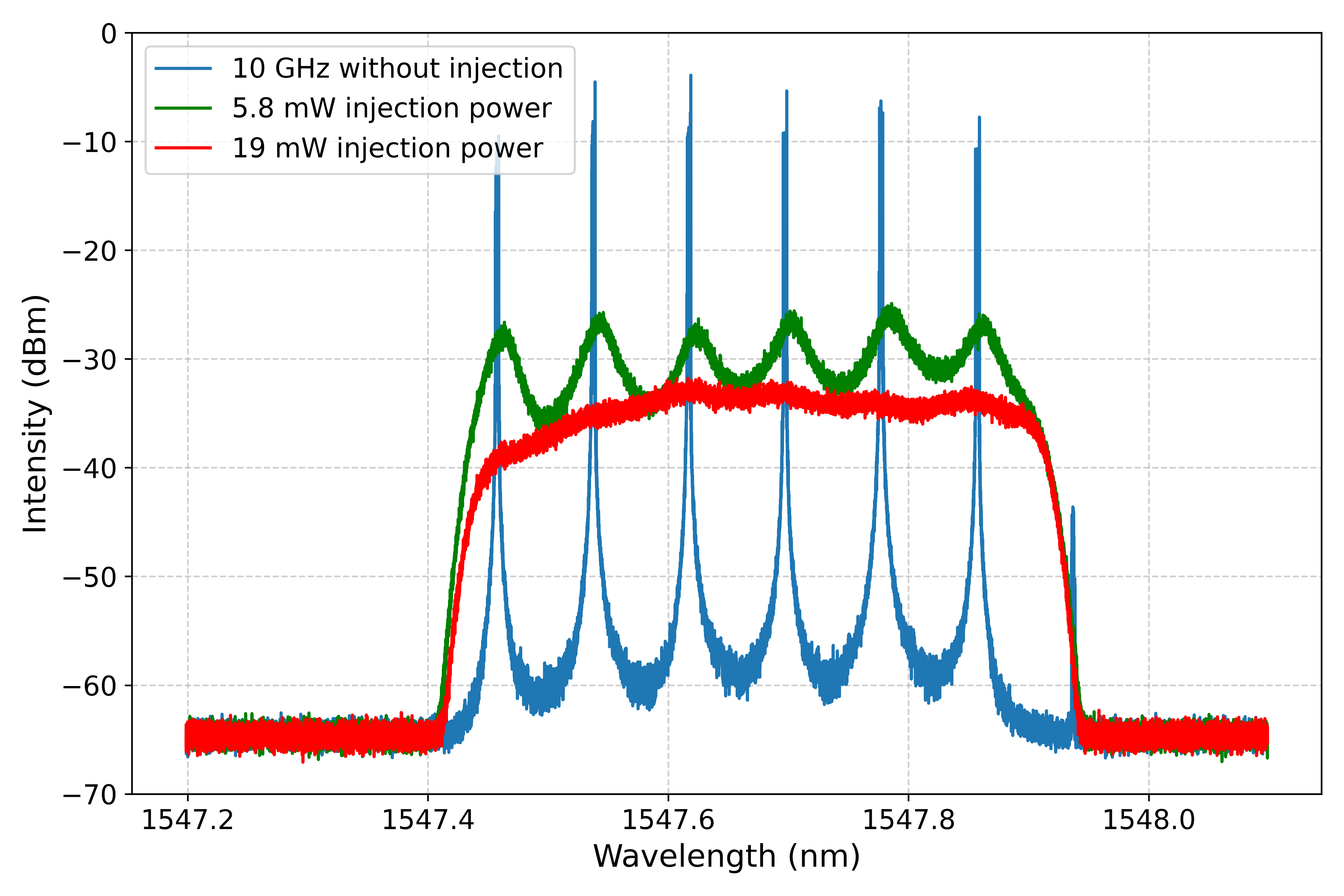}}
\caption{Optical spectra of gain-switched DFB laser at (a) 1 GHz without any injection and (b) 10 GHz under different injection powers. The high injection power of the ASE light arises from its broad spectral width of 33 nm. An optical filter centered at 1547.8 nm with a 0.5 nm bandwidth is used to filter the output.}
\label{spectrum}
\end{figure*}

The optical spectra of the laser at different modulation frequencies are measured using an optical spectrum analyser with a resolution of 1.16 pm (Fig. \ref{spectrum}). At 10 GHz without external ASE photons injection (Fig. \ref{spectrum}b), the spectrum exhibits a frequency comb with discernible tones, indicating strong pulse-to-pulse coherence. This arises from successive pulses being seeded by residual photons from previous emissions \cite{analytical}. The tone spacing is 0.08 nm, corresponding to the modulation frequency. The continuum beneath the comb lines reflects the noise in the signal periodicity, originating from pulse timing jitter induced by spontaneous emission. 

In contrast, when the laser is modulated at 1 GHz, the spectrum is continuous with no discernible tones (Fig. \ref{spectrum}a). This indicates that the phase relation between each pulse is random, as the laser is fully switched off between pulse generation and each pulse is seeded by spontaneous emission. These types of spectra have been both experimentally \cite{Rosado2018} and theoretically studied \cite{analytical}. For comparison, under optical injection locking—where external \textit{coherent} CW light seeds the gain-switched laser, well-defined frequency comb lines emerge, as pulse-to-pulse coherence is established when the laser locks to the injected light \cite{Multicarrier, analytical, Rosado2018}.

The spectral response under external ASE photons injection is shown in Fig. \ref{spectrum}b. As the injection power is increased, the comb gradually broadens due to the enhanced spontaneous emission rate, which leads to an accelerated phase diffusion. Eventually the comb structure vanishes, the resulting spectrum closely resembles that of the 1~GHz case. The presence of a smooth continuum without comb features suggests that the pulse-to-pulse coherence has been effectively suppressed by the injected ASE photons, thereby restoring the phase randomness at each pulse generation even at high clock rates. The disappearance of frequency comb structure is consistent with loss of pulse-to-pulse coherence, and provides a qualitative supporting indicator of phase randomization.

We characterized the pulse timing jitter as a function of the injected ASE power, with results summarized in Table \ref{tab:timing-jitter}. Spontaneous emission influences not only the phase of the optical pulses but also their timing jitter, owing to the uncertainty in the build-up time of positive gain in the cavity. The increase in timing jitter under ASE injection is therefore consistent with the stochastic pulse build-up process introduced by spontaneous-emission seeding. At 10 GHz, an ASE injection power of 19 mW was chosen, which increased the timing jitter from 4.4 ps to 21.9 ps. The injection power therefore represents a trade-off: if it is too low, the seeding effect is insufficient to fully randomize the phase; if it is too high, the excessive jitter prevents sufficient temporal overlap of the pulses, thereby degrading interference. At the chosen operating point, this overlap nonetheless remains sufficient, as evidenced by the recovery of the arcsine intensity distribution in Fig.~\ref{Img:Exp10g}c, confirming that the injection level lies within the usable window of this trade-off.

\begin{table}[htbp]
  \centering
  \caption{Timing jitter at different ASE injection power}
  \label{tab:timing-jitter}
  \begin{tabular}{cc}
    \hline
    \textbf{Injection power (mW)} & \textbf{Timing jitter (ps)} \\
    \hline
    0  & 4.4  \\
     5  & 15.1 \\
     19 & 21.9 \\
    24 & 28.3 \\
    \hline
  \end{tabular}
\end{table}

It is worth noting that the interference degradation referred to above concerns the interference between \textit{consecutive} pulses, as used here to characterise the phase randomness. In phase-encoded QKD, by contrast, each pulse is split into two temporal modes in Alice's interferometer and recombined in a matching interferometer at Bob, so the first-order interference that carries the key information occurs between two replicas of the \textit{same} pulse. The pulse timing jitter is then common to both interfering components and cancels, leaving the interference visibility unaffected. Instead, the jitter manifests as a constraint on the detection time-gating and inter-symbol spacing rather than on the interference itself.

\section{Discussion}
We have demonstrated a simple and robust scheme to randomize the phase of optical pulses even when the laser is modulated at high repetition rates up to 10 GHz. This is achieved by externally injecting ASE photons from an SLD into the laser cavity, which allows the generation of each pulse to be seeded by spontaneous emission photons so that each pulse acquires a random phase.  

Based on a rate-equation approach following Lovic et al.\cite{Lovic2021PhaseNoiseQRNG}, modelling indicates that the achievable repetition rate scales logarithmically with injection power. At rates around 10~GHz, the required ASE power is readily accessible; pushing substantially beyond this, higher injection powers become less effective due to the logarithmic scaling. In this regime the photon lifetime emerges as the limiting factor to realising higher clock rates. Higher clock rates are therefore most effectively reached by reducing the laser's effective photon lifetime, for example through realising a shorter or higher-loss laser diode cavity, at which point the laser's modulation bandwidth becomes the relevant consideration. As the repetition rate approaches the laser's modulation bandwidth, the carrier density can no longer be fully driven below threshold each cycle, preventing complete decay of the coherent field. A thorough and precise determination of the upper clock-rate limit is the subject of further study.

It is useful to obtain a first approximation of the generation rate of random numbers that could be achieved with this scheme. This can be estimated by evaluating the theoretical min-entropy $H_{min}(X \vert E)$ of the random variable $X$, associated with the pulse intensity (digitized by an ideal 8-bit analog-to-digital converter (ADC)), conditioned on the electronic noise $E$ \cite{Marangon2024}:

\begin{equation}
    H_{min}(X \vert E) = -\log_2 \left[1-\frac{2}{\pi} \tan^{-1}(\sqrt{\frac{\Delta_{CD}-\delta}{\delta}}) \right]
\end{equation}

\noindent where $\Delta_{CD}$ is the width of the arcsine distribution and $\delta$ is the ADC bin width. For the data acquired at 10 GHz, we obtain $H_{min}(X \vert E)$ = 4.321 bit/pulse with 99.999\% confidence, corresponding to an ideal generation rate exceeding 40 Gbit/s. This estimate does not account for digitizer non-idealities, real-time extraction overhead, or a complete entropy-source model. A full QRNG implementation with real-time digitization and randomness extraction is beyond the scope of the present work. To produce usable digital output, the raw bit strings must be compressed through post-processing algorithms. Even assuming a conservative 50\% compression factor to account for hardware non-idealities, the effective generation rate remains above 20 Gbit/s.

This approach overcomes one of the main obstacles limiting the clock rate — and therefore the key rate — of current QKD systems. While our results demonstrate the feasibility of performing high-rate discrete-variable QKD at up to 10 GHz, achieving this in practice also requires availability of other system components capable of matching such clock rate. In particular, avalanche photodiodes (APDs), the most commonly used detectors in deployed QKD systems, typically operate at 1 GHz, and APDs capable of 10 GHz operation have yet to be developed. However, superconducting nanowire single-photon detectors (SNSPDs), which usually operate in free-running mode, offer a promising alternative and are increasingly gaining popularity. Combined with a commercially available state-of-the-art timetagger with a resolution of a few picoseconds, single-photon detection at 10 GHz should be presently feasible. For signal encoding, electro-optic modulators with bandwidths up to 40 GHz are also readily available off the shelf. Therefore our result presents an exciting development on the path to demonstrating 10-GHz-clocked QKD. Naturally, a complete QKD demonstration will also require a careful assessment of the practical penalties introduced by the technique—for example the increased timing jitter, residual ASE or any remaining correlations between pulse parameters, which we leave to future work.

To conclude, we have demonstrated that interpulse phase correlation in gain-switched lasers can be overcome by externally injecting ASE photons into the laser cavity. This enables phase randomization to be achieved at repetition rate as high as 10 GHz. This approach opens a new avenue for the development of high-rate QRNGs and QKD systems.

\section*{Funding}
Toshiba Europe Limited; Engineering and Physical Sciences Research Council (210138). European Union’s Horizon Europe Framework Programme under the Marie Sk\l{}odowska Curie Grant No.101072637, Project Quantum-Safe Internet (QSI).

\section*{Author contributions}
Y.S.L. and R.I.W. conceived the idea. Y.S.L. and A.B. performed the experiments, with support from P.R.S., D.G.M., S.J. and J.F.D. A.B. performed the numerical analysis, with support from P.R.S., R.I.W., D.G.M. and T.P.. R.M.S. and A.J.S. supervised the project. All authors contributed to the writing and editing of the manuscript.

\section*{DISCLOSURES}
The authors declare no conflicts of interest.

\section*{Data Availability Statement}

The data that support the findings of this study are available from the corresponding author upon reasonable request.

\appendix

\section*{Appendix}

\begin{figure*}[tbph!]
\includegraphics[scale=0.4]{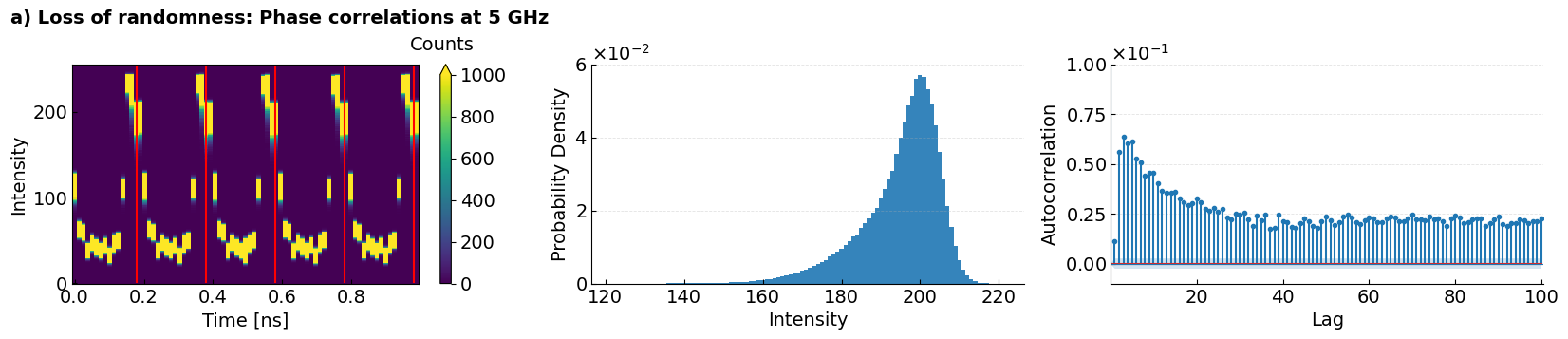}
\includegraphics[scale=0.4]{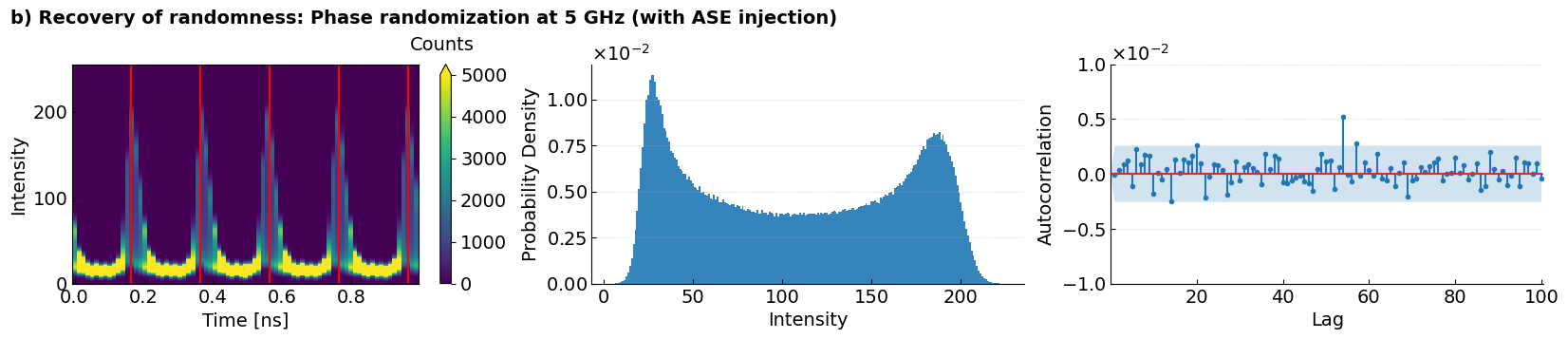}
\caption{Experimental results of the DFB laser under gain-switching at (a) 5 GHz, (b) 5 GHz with external ASE light injection. \textbf{Left column:} Density plots of the oscilloscope traces, with the sampling points indicated by red lines. \textbf{Middle column:} The corresponding intensity distributions at the sampling point. \textbf{Right column:} Autocorrelation functions of the intensity at the sampling point, computed between samples separated by a specified number of lags. The blue shaded region denotes the 99\% confidence interval bounds and lag 0 is omitted for clarity. Analyses are performed over $1 \cdot 10^6$ samples.}
\label{Img:Exp5g}
\end{figure*}

\begin{figure*}[tbph!]
\includegraphics[scale=0.4]{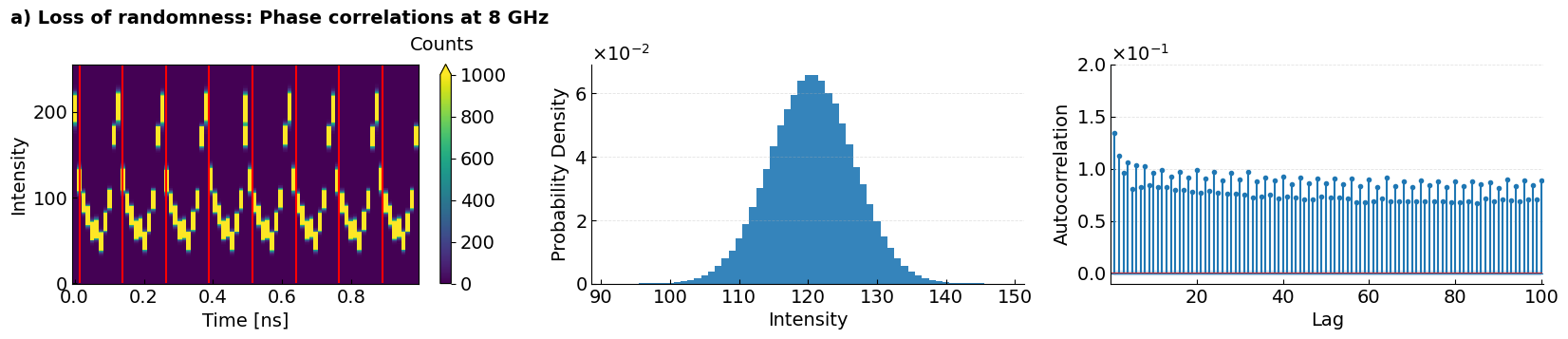}
\includegraphics[scale=0.4]{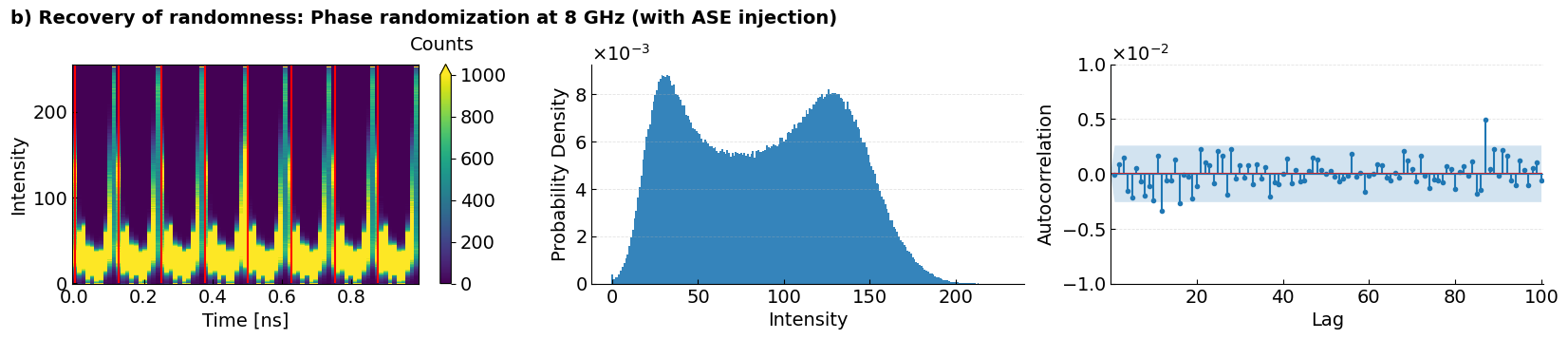}
\caption{Experimental results of the DFB laser under gain-switching at (a) 8 GHz, (b) 8 GHz with external ASE light injection. \textbf{Left column:} Density plots of the oscilloscope traces, with the sampling points indicated by red lines. \textbf{Middle column:} The corresponding intensity distributions at the sampling point. \textbf{Right column:} Autocorrelation functions of the intensity at the sampling point, computed between samples separated by a specified number of lags. The blue shaded region denotes the 99\% confidence interval bounds and lag 0 is omitted for clarity. Analyses are performed over $1 \cdot 10^6$ samples.}
\label{Img:Exp8g}
\end{figure*}

The 5 GHz result (Fig. \ref{Img:Exp5g}) was collected whilst biasing the laser at 42 mA, with the RF at 4.6 V peak-to-peak. The 8 GHz case (Fig. \ref{Img:Exp8g}) had the same RF waveform, but the laser was biased at 42.5 mA. In both cases the SLD injection power was 5.7 mW.

\bibliography{biblio}

\end{document}